\documentstyle[preprint,epsfig]{jpsj}
\newcommand{\tr}{{\rm Tr}}
\newcommand{\ham}[1]{{\cal H}_{\lambda(#1)}}
\newcommand{\work}[1]{\hat{W}(#1)}

\title{A Quantum Analogue of the Jarzynski Equality}
\author{Satoshi {\sc Yukawa}\footnote{E-mail address:
    yukawa@ap.t.u-tokyo.ac.jp}} 
\inst{Department of Applied Physics, School of Engineering,
  The University of Tokyo,\\7-3-1, Hongo, Bunkyo-ku, Tokyo 113-8656,
  JAPAN}
\recdate{May 24, 2000}
\abst{
  A quantum analogue of the Jarzynski equality is constructed. 
  This equality connects an ensemble average of exponentiated work
  with the Helmholtz free-energy difference in a nonequilibrium
  switching process subject to a thermal heat bath.
  To confirm its validity in a practical situation, 
  we also investigate an open quantum system that is a spin 1/2
  system with a scanning magnetic field interacting with a thermal heat
  bath. As a result, we find that the quantum analogue functions well.
}
\kword{quantum nonequilibrium dynamics, Jarzynski equality}

\begin{document}
\sloppy
\maketitle

Understanding nonequilibrium quantum dynamics well be an important problem
in the next century; new types of quantum devices, for instance, may
function in their nonequilibrium states. In fact, it has been
theoretically shown that
quantum ratchet systems, which have the potential for being a new type of
quantum device, function well in competitive situations between
thermal relaxation and external driving, which are extremely
nonequilibrium conditions\cite{YKTM97,RGH97,TKYM98}.
A framework of theoretical methods for understanding such systems,
however, is still under development.  In this letter, we shall develop
one equality which holds during a nonequilibrium process accompanied with
thermal relaxation and external driving.  This equality is a quantum
analogue of the Jarzynski equality\cite{Ja96,Ja97}. Considering such
equality raises important issues regarding quantum nonequilibrium
thermodynamics.

Consider the situation in which a classical system interacts with an
isothermal heat bath characterized by an inverse temperature $\beta$. In
this case, the system will become a thermal 
equilibrium state after a sufficiently long relaxation time.
Next we assume that the system
depends on a (macroscopic) parameter
$\lambda$. If we change the
parameter $\lambda$ infinitely slowly from $0$ to $1$, the system
becomes another thermal equilibrium state corresponding to $\lambda =
1$ from an equilibrium state with $\lambda = 0$. When the operation is
sufficiently slow (switching time $t_s$ is 
much larger), that is, when the system remains in quasistatic
equilibrium during a switching process, the total
{\it work} $W(t_s=\infty)$ performed on the system by the outside
through the parameter 
$\lambda $ is identical to the Helmholtz free-energy difference $\Delta
F$ between the initial state and the final one:
\[
W(t_s = \infty) = \Delta F \equiv F_0 - F_1 
\enspace ,
\]
where $F_\lambda$ is the Helmholtz free energy of the thermal equilibrium
state with $\lambda$. 
If we operate the switching much faster, that is, a finite $t_s$,
the work must be larger than the free-energy difference:
\[
W(t_s) \ge \Delta F 
\enspace .
\]
This inequality represents the least-work principle. Extra work is
required for a faster switching process because of thermodynamic
stability. This extra work is released into the heat bath
after the operation, when the system interacts with
the isothermal heat bath.

In 1997, Jarzynski presented an equality which differs from
the above inequality in the same situation\cite{Ja96,Ja97}. His result is
expressed by the 
following form: 
\begin{equation}
\langle \exp ( -\beta W(t_s) ) \rangle_{path} = \exp( -\beta \Delta F)
\enspace .
\label{eq:cljar}
\end{equation}
This is called the Jarzynski equality. 
The brackets $\langle \dots \rangle_{path}$ mean taking the 
ensemble average along a single path during 
the switching process. An initial state is chosen from a thermal
equilibrium ensemble with $\beta$ and $\lambda=0$. 
The dynamics of the system during the process can be taken as an 
arbitrary one that can express the thermal relaxation. If Monte Carlo
dynamics, for instance, is used, one path corresponds to a single
sampling series. Jarzynski also showed that the above equality  holds
in several 
dynamics such as Monte Carlo dynamics, Nos\'e-Hoover dynamics,
Langevin dynamics, and even Hamiltonian dynamics without a thermal
heat bath\cite{Ja97}. 
This equality is general since it involves the least-work principle
mentioned above.

In this letter, we consider an analogue of the Jarzynski equality in quantum
situations. Let us define several notations and quantities. 
We denote the Hamiltonian of the system as $\ham{t}$, and
$\lambda(t) $ a switching parameter operated by the outside at time
$t$.   A partition
function $Z_\lambda$ and the Helmholtz free energy $F_\lambda $ are
expressed as follows:
\begin{eqnarray}
  Z_{\lambda(t)} & = \tr \exp ( -\beta \ham{t} ) \enspace, \\
  F_{\lambda(t)} & = - \beta^{-1} \ln Z_{\lambda(t)}(\beta) \enspace .
\end{eqnarray}
The work operator $\hat{W}(t)$ during a time interval $[0,t]$ is defined
by the following: 
\begin{equation}
\work{t} \equiv \int_0^{t} ds \dfrac{\partial \lambda(s)}{\partial
  s} \dfrac{\partial \ham{s}}{\partial \lambda}  \enspace .
\label{eq:workop}
\end{equation}
If $\lambda$ is not a macroscopic parameter, this definition does not
correspond to the intuitive view of work. Here, we only consider the
case where $\lambda$ is a macroscopic parameter. 
Using a density matrix of the system $\rho(t)$, we get the
expression of the total average work $\langle \work{t} \rangle$ in the
switching process as  
\begin{equation}
  \langle \work{t} \rangle \equiv \tr \left\{
    \int_0^t ds \dfrac{\partial \lambda(s)}{\partial
      s} \dfrac{\partial \ham{s}}{\partial \lambda}  \rho(s)
  \right\} 
  \enspace .
  \label{eq:workave}
\end{equation}
In the present situation, the least-work
principle also holds
\[
\langle \work{t} \rangle \ge \Delta F
\enspace 
\]
for arbitrary switching processes. 
The equal sign is valid for quasistatic limits as 
well as the classical case.

Let us consider a path average of exponentiated work corresponding
to the left-hand side of the classical Jarzynski equality. 
In the present situation, taking the path average is difficult because
we do not know how an ensemble reproduces thermal relaxation in the
quantum dynamics. 
Therefore, we use the density matrix approach here. Then, the 
classical path average can be directly interpreted into the quantum situation:
\begin{equation}
\langle \exp \left( -\beta W(t) \right) \rangle_{path} 
\approx
\overline{
\exp \left( -\beta \work{t} \right) 
}
\equiv
\tr \left\{ 
  \lim_{N \to \infty} 
  {\cal T} \prod_{n=0}^{N-1}
  \left\{ 
    \widetilde{{\cal P}}_{\lambda(t_{n+1})}^{\delta t}
    e^{ - \beta \ham{t_{n+1}} } 
    e^{ + \beta \ham{t_{n}}   }
  \right\}  
  \frac{e^{-\beta \ham{0}}}{Z_{\lambda(0)}}
  \right\}
  \enspace ,
  \label{eq:pathavg}
\end{equation}
where $\delta t = t/N$ and $t_n = n \delta t$. The term $e^{ - \beta
  \ham{t_{n+1}} }  
e^{ + \beta \ham{t_{n}}}$ corresponds to exponentiated
infinitesimal work.   
In this expression,
$\widetilde{{\cal P}}_{\lambda(t)}^{\delta t}$ is
an infinitesimal time-evolution superoperator for a tiny step $\delta
t$ at time $t$ defined by
\[
\widetilde{{\cal P}}_{\lambda(t)}^{\delta t} =
e^{{\cal L}(t)\delta t} \enspace,
\]
where ${\cal L}(t)$ is the Liouville superoperator governing the time
evolution of the density matrix: 
\[
\frac{\partial \rho(t)}{\partial t} = {\cal L}(t) \rho(t)
\enspace .
\]
Details of the dynamics are not important at this time. Its least
property is determined in the following.
${\cal T}$
shows that the time ordering is taken as an increase in the left
direction.

This is actually the analogue of the classical path average. It can be
expressed as follows: 
\begin{equation}
\langle \exp ( -\beta W(t) ) \rangle_{path} 
= \lim_{ N \to \infty } \int dx_N \prod_{n=0}^{N-1} 
\left\{ 
  \int dx_n
  {\cal P}_{\lambda(t_{n+1})}^{\delta t} 
  (x_{n+1} | x_{n} ) 
  e^{-\beta \delta w(x_n) } 
\right\} 
\frac{e^{-\beta \ham{t_0}(x_0)}}{Z_{\lambda(0)}}
\label{eq:classicalave}
\enspace ,
\end{equation}
where $x_n$ denotes canonical variables describing the classical system
and an infinitesimal work, $\delta w(x_n) $ is defined by $
  \ham{t_{n+1}}(x_n) - \ham{t_{n}}(x_n) $. 
${\cal P}_{\lambda(t_{n+1})}^{\delta t} (x_{n+1} | x_{n} )$ is an
infinitesimal time-evolution operator such as a transition probability
for the Monte Carlo dynamics. From this form we can understand that
eq.~(\ref{eq:pathavg}) is a straight extension of classical
situations. 

In this way, we obtain the quantum analogue of the path-averaged
exponentiated work. In the next step, we must show that the above
expression is identical to the exponentiated Helmholtz free-energy
difference. For this purpose, we use the fundamental property of the
Liouville superoperator: it is vanished by the one density matrix that
is the thermal equilibrium distribution $\propto \exp \left( -\beta
  \ham{t} \right)$, because the system always interacts with the
thermal heat bath in the present case. 
Such distribution, of course, depends on time $t$. That is, $\exp
\left( -\beta \ham{t} \right)$ is a singular solution of   
the dynamics.
In this sense, a part of the exponentiated infinitesimal work does
not evolve: 
\[
\widetilde{{\cal P}}_{\lambda(t_{n+1})}^{\delta t}
e^{ - \beta \ham{t_{n+1}} } 
= e^{ - \beta \ham{t_{n+1}} } 
\enspace .
\]
Finally, we obtain
\begin{eqnarray*}
&\tr & \left\{ 
  \lim_{N \to \infty} 
  {\cal T} \prod_{n=0}^{N-1}  
  \left\{ 
    \widetilde{{\cal P}}_{\lambda(t_{n+1})}^{\delta t}
    e^{ - \beta \ham{t_{n+1}} } 
    e^{ + \beta \ham{t_{n}}   }
  \right\}  
  \frac{e^{-\beta \ham{0}}}{Z_{\lambda(0)}}
  \right\}\\
& = &
  \frac{1}{Z_{\lambda(0)}}\tr \left\{ 
  \lim_{N \to \infty} 
  \widetilde{{\cal P}}_{\lambda(t_{N})}^{\delta t}
  e^{ - \beta \ham{t_{N}}    } 
  e^{ + \beta \ham{t_{N-1}}   }
  \dots
  \widetilde{{\cal P}}_{\lambda(t_1)}^{\delta t}
  e^{ - \beta \ham{t_1} } 
  e^{ + \beta \ham{0}   }
  e^{-\beta \ham{0}}
  \right\}\\
& = &
  \frac{1}{Z_{\lambda(0)}}\tr \left\{ 
  \lim_{N \to \infty} 
  \widetilde{{\cal P}}_{\lambda(t_{N})}^{\delta t}
  e^{ - \beta \ham{t_{N}}    } 
  e^{ + \beta \ham{t_{N-1}}   }
  \dots
  \widetilde{{\cal P}}_{\lambda(t_2)}^{\delta t}
  e^{ - \beta \ham{t_2} } 
  e^{ + \beta \ham{t_1}   }
  e^{-\beta \ham{t_1}}
  \right\}\\
&\vdots&\\
&=&
  \frac{1}{Z_{\lambda(0)}}\tr \left\{
  e^{-\beta \ham{t}}
\right\} 
\enspace .
\end{eqnarray*}
Thus, the following equality
\[
  \overline{\exp\left( -\beta \work{t} \right) } 
  = 
  \frac{Z_{\lambda(t)}}{Z_{\lambda(0)}}
\]
is proofed. This is the quantum analogue of the classical Jarzynski
equality, the ``quantum Jarzynski equality.''

It should be noted that this expression is valid in pure quantum
dynamics as in classical dynamics. In pure dynamics, an operation of
the superoperator on an arbitrary operator $A$,
$\widetilde{{\cal P}}_{\lambda(t_{n+1})}^{\delta t} A$, is
identical 
to $ e^{-i \ham{t_{n+1}} \delta t /\hbar} A e^{i \ham{t_{n+1}} 
  \delta t /\hbar}$. 
Then, all the infinitesimal time-evolution superoperators are effectively
identity operators.  Finally, we obtain the exponentiated
free-energy difference again.

The quantum Jarzynski equality is formulated by an operator form. Thus,
all the variables during the process are operators in contrast to the
classical one. This property creates some difficulties in the
decomposition of the time variable. If we change the
definition of the infinitesimal exponentiated work slightly,
for example, to $e^{+\beta\ham{t_n}} e^{-\beta\ham{t_{n+1}}}$
or $ e^{-\beta (\ham{t_{n+1}}-\ham{t_n})}$, it causes errors at each
time. Denoting the error of order $O(\delta t)$ as $A_{t_n}$, that is,
the exponentiated work is taken as $e^{-\beta\ham{t_{n+1}}}
e^{+\beta\ham{t_n}} + A_{t_n} + O((\delta t)^2)$, we obtain the final
error as 
\begin{equation}
\dfrac{1}{Z_{\lambda(0)}}
\tr 
\left[
  \lim_{ N \to \infty}
  \left\{
    \sum_{l=0}^{N-1} 
    \left(
    {\cal T}{\prod_{n=l+1}^{N-1}}^*
    \widetilde{{\cal P}}_{\lambda(t_{n+1})}^{\delta t}
    e^{ - \beta \ham{t_{n+1}} } 
    e^{ + \beta \ham{t_{n}}   }
    \right)
    \widetilde{{\cal P}}_{\lambda(t_{l+1})}^{\delta t}
    A_{t_l}
    e^{ -\beta \ham{t_l}}
    + O((\delta t)^2)
  \right\} 
\right]
\enspace,
\end{equation}
where ${\prod_{n=l+1}^{N-1}}^*$ represents the omission of 
the term $\prod_{n=N}^{N-1}$. 
At the limit $N \to \infty$ this error may remain, since $N$ terms of
the order $O(\delta t)=O(1/N)$ are collected. The higher-order errors may
vanish because their number is $O(N)$ at most. Therefore, the small
mistaken choice of the exponentiated infinitesimal work causes a
serious error in the final result. In a practical situation, however, 
$A_{t_n}$ may be of the order $O(\delta t \times (t_s)^{-1})$. For
sufficiently large $t_s$, the final error is not serious.

To confirm the quantum Jarzynski equality, we numerically investigate
a simple quantum system. We choose the spin $1/2$ system 
interacting with an isothermal heat bath and time-dependent magnetic
field. Here, we consider a process in which the magnetic field is linearly
reversed in a finite time interval.
The Hamiltonian ${\cal H}_s(t)$ and the magnetic field
$\lambda(t)$ at time $t$ are as follows:
\begin{eqnarray*}
{\cal H}_s (t) & = - \frac{\lambda(t)}{2} \sigma^z - \frac{1}{2}
\Delta \sigma^x \enspace , \\
\lambda(t) & = \lambda_0 \left( 1-2 \frac{t}{t_s} \right)
\enspace ,
\end{eqnarray*}
where $\sigma^x$ and $\sigma^z$ are Pauli matrices. The parameters
$\Delta$ and $\lambda_0$ are constants taken as $\Delta = 0.1 $ and $
\lambda_0 = 1$, respectively. 
Interaction between the system and the heat bath is described by 
$\gamma \sigma^x \sum_\alpha (a^\dagger_\alpha + a_\alpha)$ ($\gamma$
is a coupling constant) and the bath is expressed by a set of harmonic
oscillators $ \sum_{\alpha} \hbar \omega_\alpha \left(
  a^\dagger_\alpha a_\alpha + 1/2 \right) $ , where $a^\dagger_\alpha
(a_\alpha)$ is a creation (annihilation) operator with a mode
$\alpha$. $t_s$ is a switching interval. The dynamics starts at
time $t=0$ with a thermal equilibrium state and finishes at time
$t=t_s$.  
Using the projection technique\cite{KTH91} with the assumption that the 
heat bath is always in a thermal equilibrium state with an inverse
temperature $\beta$, we obtain the equation of motion of the density
matrix of the spin system $\rho(t)$ as 
\begin{equation}
\frac{\partial \rho}{\partial t} = - \frac{i}{\hbar} \left[
{\cal H}_s(t) , \rho(t) 
\right] + \frac{\gamma^2}{\hbar^2} \Gamma(\rho)
\enspace ,
\label{eq:spindyna}
\end{equation}
where $\Gamma$ is a relaxation term guaranteeing that the system will relax
into an instantaneous thermal equilibrium state $ \propto \exp ( -\beta
{\cal H}_s(t) )$. If the Hamiltonian does not depend on time, this
equation can express the thermal relaxation to a thermal equilibrium state.
The explicit form of $\Gamma$ is so complicated that is not
presented here, but can be essentially described by $\sigma^x$ (interaction
between the spin and the bath) and the
autocorrelation function 
of the thermal heat bath variables. For actual calculations we use the
second-order perturbation  for the coupling constant $\gamma$.

\begin{figure}
\epsfig{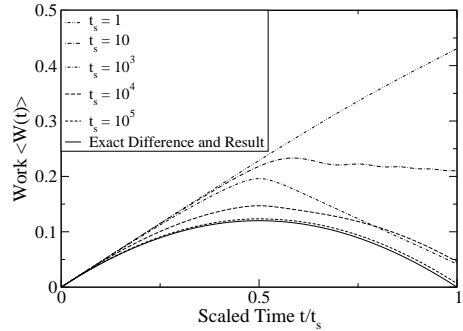}
\caption{Time series of the expectation value of work, the exact
  Helmholtz free-energy difference and the result of the quantum
  Jarzynski equality. Horizontal axis is taken to be scaled time
  $t/t_s$.}
\label{fig:spin}
\end{figure}

Figure~\ref{fig:spin} shows the results of numerical calculations. Several
time series of the expectation value of work defined by
eq.~(\ref{eq:workave}) are presented. Each time series has a different
switching time $t_s$. 
The exact Helmholtz free-energy
difference at the corresponding time and the result of the quantum
Jarzynski equality are also shown. As switching time increases, the
value of work is more converged into the exact Helmholtz free-energy 
difference, because the dynamics becomes closer to a quasistatic one. On
the other hand, when $t_s$ is short, a nonadiabatic transition dominates
the dynamics. Thus, the extra work is needed.  The result of the
quantum Jarzynski equality completely agrees with the exact Helmholtz
free-energy difference.

To summarize, we have constructed a quantum analogue of the
Jarzynski equality.  This equality connects a type of average of the
exponentiated work operator during the switching process with the exact
Helmholtz free-energy difference between an initial thermal equilibrium
state and the final thermal equilibrium state, even though the actual
final state of the process is not thermal equilibrium.  We have also 
confirmed that the present equality works in practical
calculations. In the spin 1/2 system with varying magnetic field
interacting with the thermal heat bath, the result coincides with the exact
difference of the Helmholtz free energy.

The present quantum Jarzynski equality includes the classical
quality. In the classical limit, the density operator becomes a
diagonal matrix, that is, the distribution function of the corresponding classical
system. In this limit, the infinitesimal time-evolution operator
describes only the transition between diagonal elements such as 
one of a Fokker-Planck operator. Then, the average part of the quantum
one is similar to eq.~(\ref{eq:classicalave}).

It should be noted that the dynamics during the switching process can be
chosen arbitrarily as one which has only the property that the
time-dependent thermal equilibrium state $\propto \exp( -\beta
\ham{t})$ is a singular solution of the dynamics. 
This property does not restrict the dynamics to Liouvillian
dynamics with the projection technique such as eq.~(\ref{eq:spindyna}).
Even though the singular solution $ \exp( -\beta \ham{t})$ is
unstable, the dynamics having such a singular solution can produce the
same results. In this situation, however, other results derived from
the dynamics are not physical.

It is known that the classical Jarzynski equality is related to the
fluctuation theorem\cite{Cr99,GC95,GC95-2}, which is the relation of 
a distribution function of the entropy production rate, 
and is also valid in dynamics between
nonequilibrium steady states\cite{Ha99}. 
Based on these findings, we can put forth several questions. 
What is a quantum  analogue of the 
classical fluctuation theorem and how is the quantum Jarzynski equality
extended into nonequilibrium steady dynamics? These problems are now
in progress.

The author is grateful to N. Ito, S. Miyashita, and K. Saito for
valuable discussions. This work was partly supported by Grants-in-Aid
from the Ministry of Education, Science, Sports and Culture (No. 11740222).

\end{document}